\documentclass[]{interact}

\usepackage{epstopdf}
\usepackage{subfigure}

\usepackage[numbers,sort&compress,merge]{natbib}
\bibpunct[, ]{(}{)}{,}{n}{,}{,}
\renewcommand\bibfont{\fontsize{10}{12}\selectfont}
\renewcommand\citenumfont[1]{\textit{#1}}
\renewcommand\bibnumfmt[1]{(#1)}

\theoremstyle{plain}

\theoremstyle{definition}

\theoremstyle{remark}

\begin{document}
\articletype{ARTICLE TEMPLATE}
\title{Efficient STIRAP-like scheme for coherent population transfer by 
revisited optimal control theory}
\author{
\name{Amine Jaouadi\textsuperscript{a}\thanks{CONTACT Amine Jaouadi. Email: 
ajaouadi@qf.org.qa} and Mamadou Ndong\textsuperscript{b,c}}
\affil{\textsuperscript{a}Qatar Foundation, Tornado Tower, Floor 5, PO. Box 
5825, Doha, Qatar; \textsuperscript{b}Laboratoire de Physique de l'Universit\'e 
de Bourgogne, UMR 5027 CNRS et Universit\'e de Bourgogne, BP 47870, 21078 
Dijon, France; \textsuperscript{c}Coll\`{e}ge Evariste 
Galois, Unit\'e d'enseignement secondaire, 13 Rue Jean Giraudoux, 95200 
Sarcelles, France}}
\maketitle
\begin{abstract}
We demonstrate that Optimal Control Theory (OCT) with a state-dependent 
constraint which depends on the state of the system at each instant can 
reproduce the famous counterintuitive mechanism of Stimulated Raman adiabatic 
passage (STIRAP). We examine this behavior in a $\Lambda$-type three-level 
system and we show that could be applied for sequentially coupled many-level 
systems. We study the robustness of the two methods with respect to pulse 
fluctuations and the decays. We show that new OCT formulation appears to be 
more robust than STIRAP when a perturbation is introduced in the pulses. Such 
method is of great use for systems involving coherence loss such as molecular 
systems with dissociation or ionization limits. It also may find potential 
applications in the control of chemical reactions, quantum optics, and quantum 
information processing.
\end{abstract}
\section{Introduction}
The problems of coherent control and efficient population transfer between 
atomic or molecular levels have been given plenty of attention and prominence 
both theoretically and experimentally \cite{Marlan,Bergmann,Bergmann2,Vitanov, 
Fewell,Fleischhauer,Kobrak,Band,Elk,Schiemann,Jaouadi} over the past few 
decades. 
They represent a great interest for a number of applications such as quantum 
computing, optical control of chemical reactions, spectroscopy and collision 
dynamics.   
STIRAP \cite{Bergmann, Kuklinski} has proven to be a robust technique to ensure 
an almost complete level-to-level
population transfer. This transfer scheme results from a mechanism 
involving a counterintuitive sequence of two laser pulses, 
conventionally called the pump and the Stokes  pulses. The counterintuitive 
behavior is translated by an unforeseen phenomena that the Stokes pulse, 
which is applied in a second time, precedes and overlaps the pump pulse. 
  
Many studies have been carried out on the processes of STIRAP \cite{Bergmann2, 
Vitanov, 
Guerin}. According to the literature, STIRAP-type solution from Local 
Control Theory (LCT) has been already demonstrated \cite{Tannor, Bartana}. 
Compared to LCT, the optimal control theory (OCT) appears to be  more flexible 
and more efficient, particularly, for complex 
systems with many degrees of freedom. In fact, OCT requires information only 
about the initial and the final (target) states. Hence, to find the optimal 
field leading to the desired objective, OCT adopts the most appropriate route 
using a forward-backward iteration process. However, LCT needs information at 
every instant during the optimization cycle in order to ensure a monotonic 
increment in the sought objective.
By using OCT, Yuan et al. \cite{Yuan} have proposed an 
analytical derivation of a  STIRAP-type solution for specific many-level systems 
in a peculiar framework, 
by assuming that the Rabi frequency is not 
bounded.
The \textquotedblleft\textit{new}\textquotedblright variant of OCT formulated
with state-dependent constraint \cite{Koch} has 
been used by M\"uller et al. \cite{Muller} for the purpose to simulate quantum 
gates within 
polar molecules and neutral atoms systems. By  coincidence, the authors have 
observed a
STIRAP-like solution when analyzing the optimized fields.

We show in this paper that by using a 
state-dependent constraint, the counterintuitive scheme 
generated from STIRAP can indeed be achieved. 
We compare here the efficiency of 
STIRAP and the new OCT by analyzing the robustness with respect to the decay 
and 
with respect to the pulse fluctuations. Finally, we demonstrate that OCT with 
state-dependent constraint can be extended to many-level systems.  

The paper is organized as follows. In section 2, we give a brief overview of 
STIRAP and the results obtained from the $\Lambda$-type three-level system. In 
section 3, 
we first, illustrate that standard OCT can not reproduce the  counterintuitive 
scheme. Secondly, we demonstrate that the new OCT with state-dependent 
constraint leads in an automatic fashion to such a mechanism. 
Then, we compare the robustness of the new OCT method against STIRAP with 
respect to the decay and to pulses fluctuations. Before concluding we show that 
the new OCT method can be extended for multilevel systems. And finally we 
conclude in section 4.
\section{STIRAP}
\begin{figure}[h!]
\centering
\includegraphics[width=0.7\linewidth]{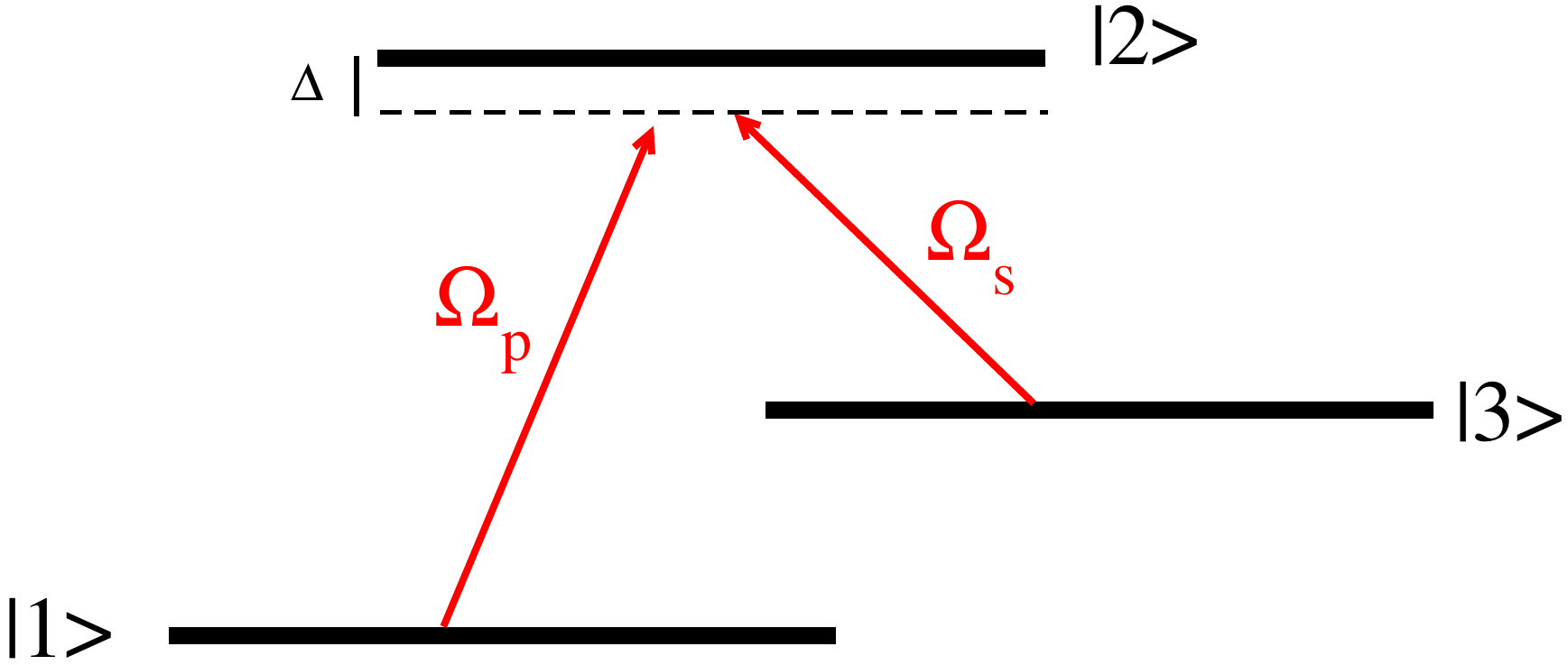}
 \caption{Schematic energy level diagram of the $\Lambda$-type three-level 
system. Levels $|1\rangle$ and $|2\rangle$ are coupled by the pump laser 
$\Omega_p$. 
Levels $|2\rangle$ and $|3\rangle$ are coupled by 
the Stokes laser  $\Omega_s$. $\Delta$ represents the detuning of 
the intermediate level $|2\rangle$. \label{fig:lambda_system}}
\end{figure}

We consider the interaction of two laser pulses with the $\Lambda$-type 
three-level system shown in Fig.~\ref{fig:lambda_system}. The levels are
sequentially coupled two-by-two: the levels 
$|1\rangle$ and $|2\rangle$ are coupled by the first field  $\Omega_p$, 
which conventionally we call the pump pulse and the levels $|2\rangle$ and 
$|3\rangle$ 
are coupled by the second field  $\Omega_s$ usually called the Stokes. The 
transition between levels $|1\rangle$ and $|3\rangle$ is electric-dipole 
forbidden. $\Delta$ represents the detuning of the intermediate level 
$|2\rangle$. We suppose as an initial condition that only level $|1\rangle$ is 
populated and the durations of the pump and the Stokes are shorter than 
the relaxation times of the system. 
We remind here that the ultimate goal of STIRAP is to transfer, in 
a efficient way, all the population from the initial state $|1\rangle$ to the 
final state 
$|3\rangle$ with minimum loss in 
the intermediate level $|2\rangle$. This has been studied intensively over the 
past few decades. However, to achieve this goal STIRAP has to go 
through a peculiar mechanism, which is manifested by the counterintuitive 
sequence of the two laser pulses. The Stokes pulse arrives before the pump 
pulse although initially we apply the pump pulse before the Stokes.
The dynamics of the three-level system are described by the 
following time-dependent Schr\"{o}dinger equation: 
\begin{equation}
 i\frac{d}{d t}a(t) = \hat{H}(t) a(t). 
 \label{eq:dynamique}
\end{equation}
Where $a(t)=[a_1(t), a_2(t), a_3(t)]^T$, $a_1(t)$, $a_2(t)$ and $a_3(t)$ are 
respectively the probability amplitudes of the states $|1\rangle$, 
$|2\rangle$, and $|3\rangle$.\ 

Using the rotating-wave approximation, the time-dependent Hamiltonian 
$\hat{H}$ of the system can simply be written as:
\begin{eqnarray}
     \hat{H}(t) &=& 
 \begin{bmatrix}
        0 &  \Omega_p(t) &  0 \\
       \Omega_p(t) & \Delta(t) & \Omega_s(t) \\
       0 &  \Omega_s(t) &  0 
 \end{bmatrix}.
 \label{eq:Ham1}
\end{eqnarray}

Where $\Omega_p(t)$ and $\Omega_s(t)$ represent the Rabi frequencies of 
the two pulses. Indexes $p$ and $s$ refer to the pump and the 
Stokes pulses respectively. 

\begin{figure}[h!]
\centering
\includegraphics[width=0.7\linewidth]{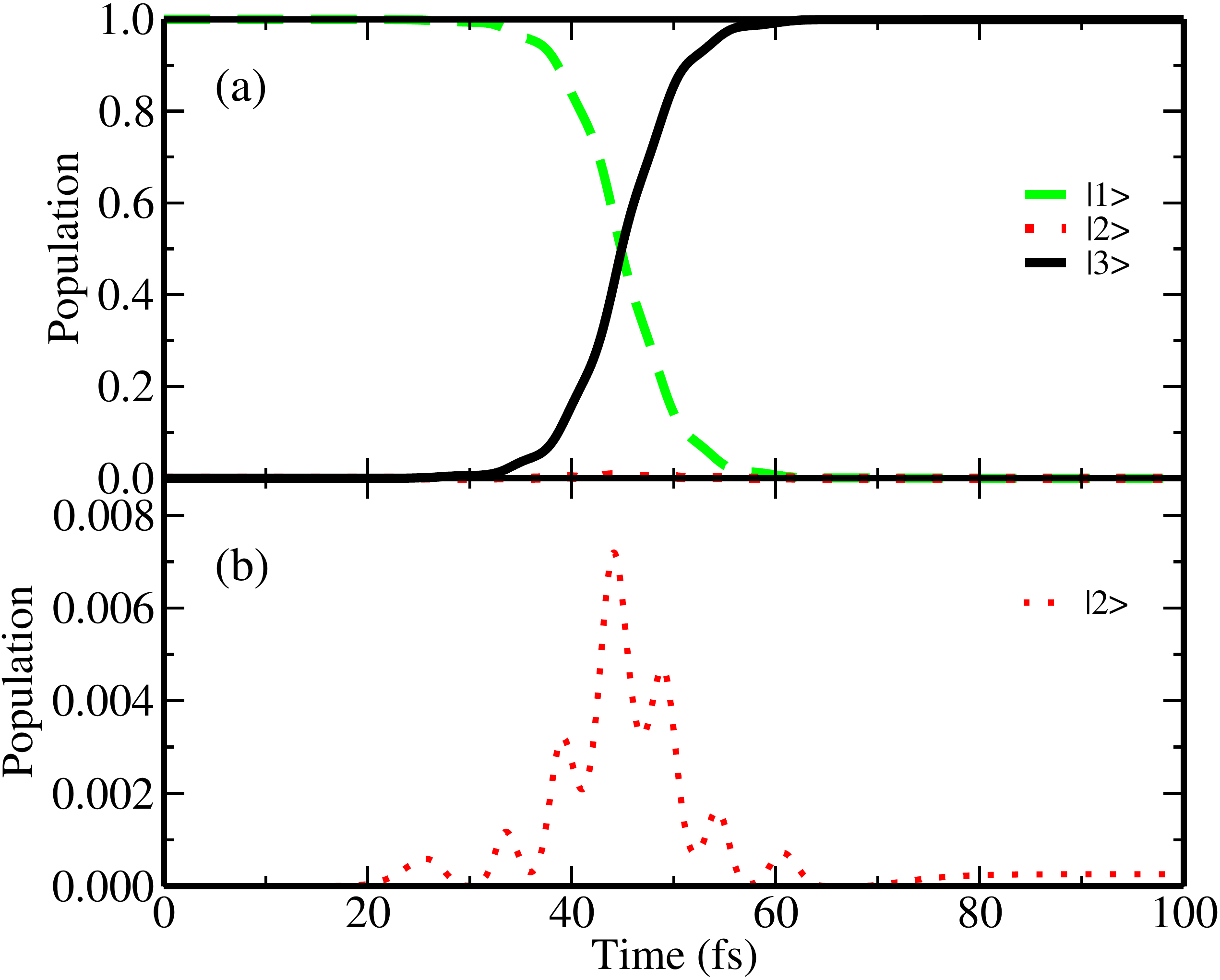}
\caption{The time evolution of the population in the three states with laser 
parameters generated by STIRAP. Level $|1\rangle$ is presented in green dashed 
line, level $|2\rangle$ in red dotted line, and level $|3\rangle$ in solid dark 
line. \label{fig:Populations-STIRAP}}
\end{figure}

\begin{figure}[h!]
\centering
\includegraphics[width=0.7\linewidth]{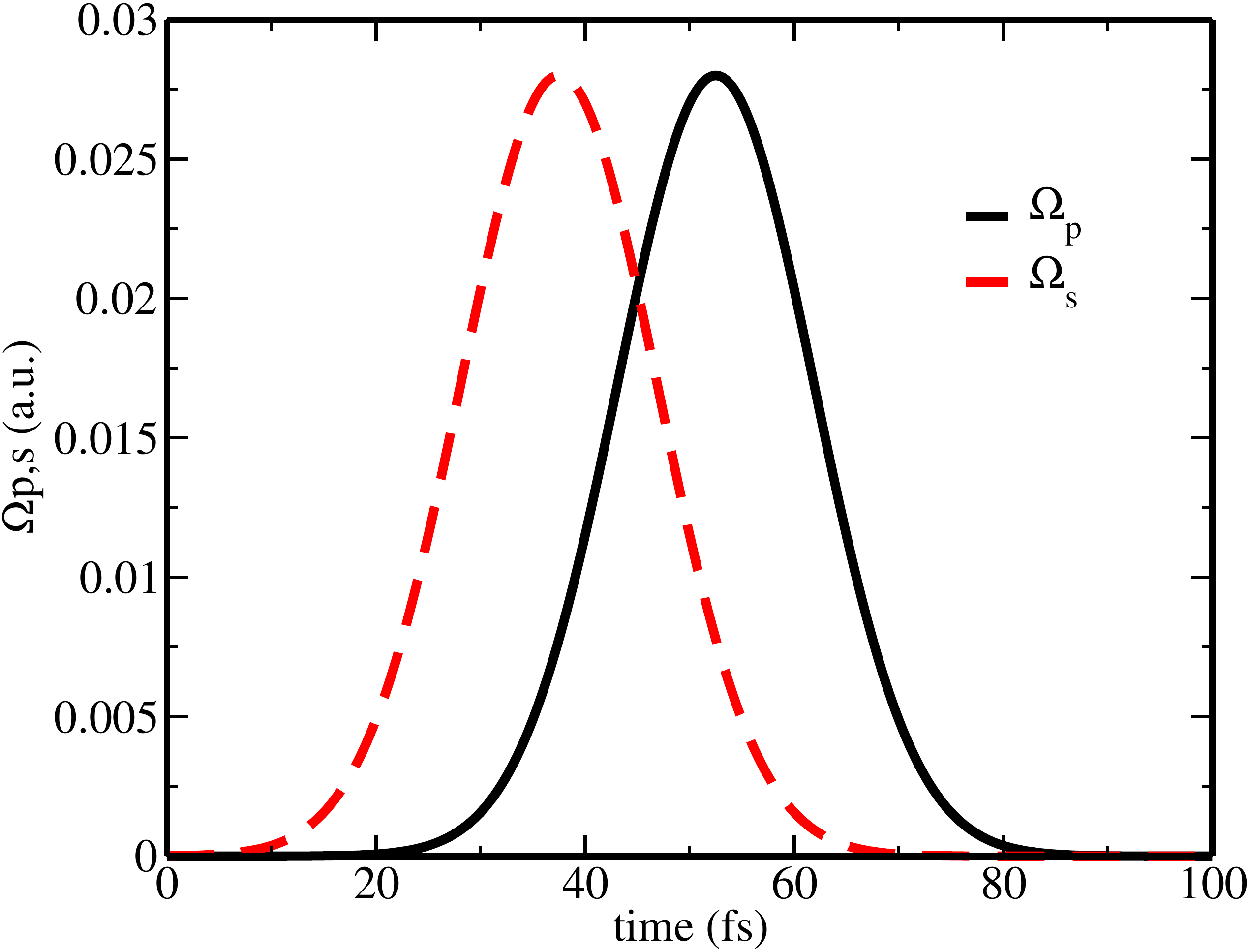}
\caption{The sequence of the two pulses produced by using STIRAP method. We 
note that the counterintuitive sequence 
is generated automatically; the Stokes pulse (dashed red line) precedes the 
pump pulse (solid dark line).
\label{fig:Pulses-STIRAP}}
\end{figure}

In Fig.~\ref{fig:Populations-STIRAP}, we display the dynamics of the 
populations as a function of time. We choose a period $T=$ 100 fs as a 
duration for both pulses. 
At time $t=T$,  STIRAP leads to an almost a complete 
transfer of population from state $|1\rangle$ (the dashed green curve) to state 
$|3\rangle$ (the solid dark curve). However, as one can notice 
in Fig.~\ref{fig:Populations-STIRAP}(a) level $|2\rangle$  (the dotted red 
curve) is not completely dark. 
For comparative purposes with other methods presented in the following section, 
we analyze in detail the population being remained in the intermediate level 
$|2\rangle$.  In Fig.~\ref{fig:Populations-STIRAP}(b), we show a zoom of the 
population of the state $|2\rangle$ as a function of time (red dotted curve). 
We observe that the population has to go through a peak at around $t\backsimeq 
45$ fs where it reaches a maximum of $\backsimeq0.8\%$. 
It is well known that STIRAP generates automatically a counterintuitive 
and unanticipated behavior, where we found once the process is over, the 
Stokes pulse is followed by the pump pulse. 
Again for purpose of comparison, we plot in Fig.~\ref{fig:Pulses-STIRAP} the 
pulses generated by 
STIRAP for a pulse duration $T=$100 fs. 
We can observe that the 
counterintuitive sequence of Stokes pulse 
(in dashed red curve) followed by the pump pulse (solid black curve) emerges 
clearly as it was expected \cite{Vitanov}.
\section{Optimal Control Theory}
In this section we first investigate the standard OCT method applied to the 
three-level $\Lambda$ system. Where the aim is the same: ensure to transfer the 
whole  population from the initial state $|1\rangle$ to the target state 
$|3\rangle$ without or with minimum of dissipation through the intermediate 
state $|2\rangle$. 
Subsequently, we analyze the implementation of state-dependent constraint 
developed by Palao et al. \cite{Koch}  in order to achieve efficient and 
complete 
population transfer. We then, study the robustness of the new OCT
method and compare it to STIRAP performance. Finally, we show that the new OCT 
with the state-dependent constraint 
can be extended for systems involving more than three levels.
\subsection{OCT Standard Method}
Generally, in the implementation of an OCT scheme, as a general rule, one 
would like to drive the system from an initial state to a specified target 
state 
at the final time. Usually the optimization 
requires to define an objective functional $J$, which  has to be 
maximized or minimized \cite{Rabitz,Vivie}. 

Here, we first formulate briefly the basics of optimal control theory for 
the $\Lambda$-type three-level system. The objective is to find the optimal 
laser field that drives the system from the initial state 
$|\varphi_i\rangle=|1\rangle$ to the target state $|\varphi_f\rangle=|3\rangle$ 
at the final time $T$. 

\begin{figure}[h!]
\centering
\includegraphics[width=0.7\linewidth]{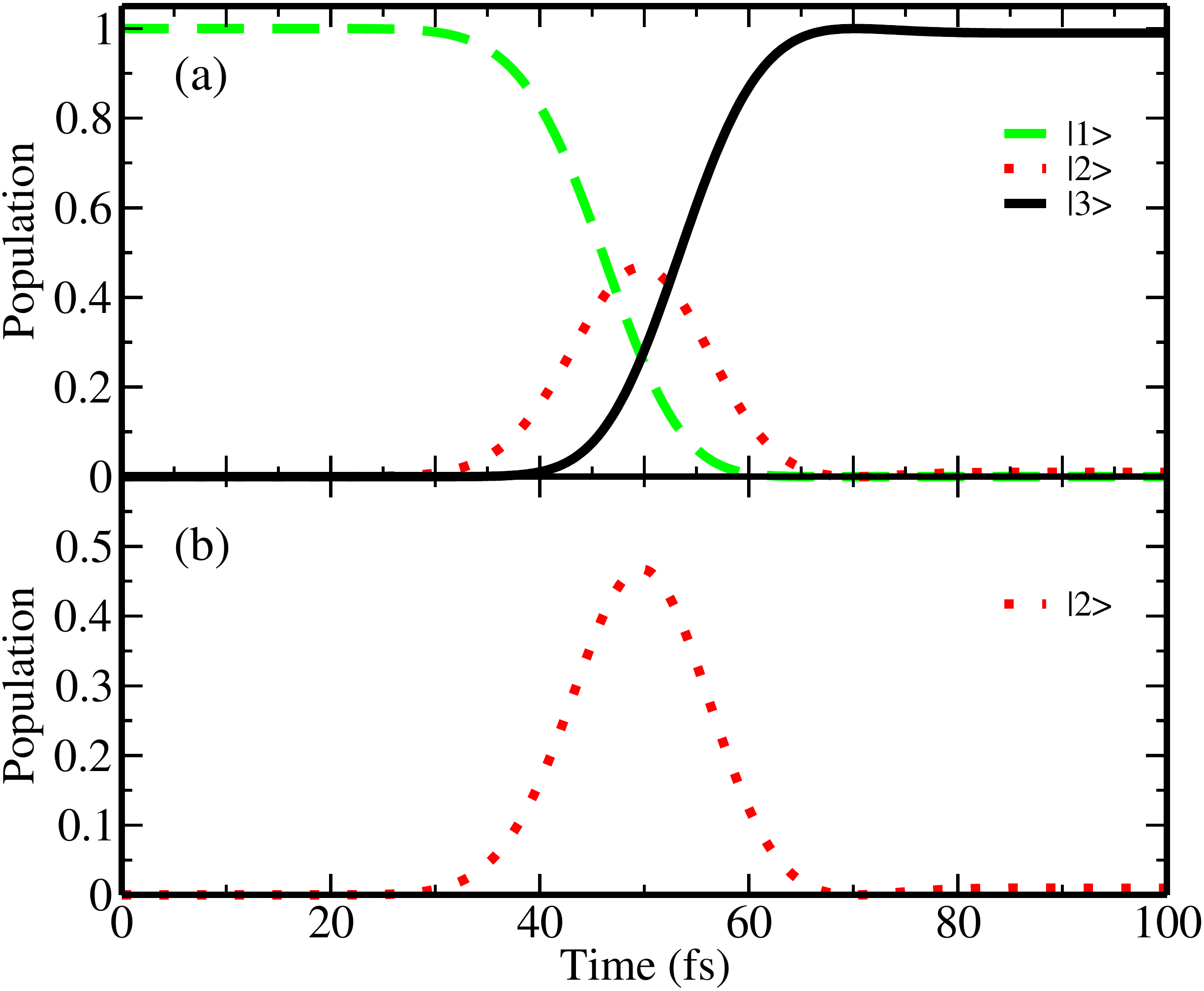}
\caption{The time evolution of the population in the three states $|1\rangle$ 
(green dashed line), 
$|2\rangle$ (dotted red line), and $|3\rangle$ (solid black line) with laser 
parameters generated by the standard OCT method.
\label{fig:Populations-OCT-Standard}}
\end{figure}

\begin{figure}[h!] 
\centering
\includegraphics[width=0.7\linewidth]{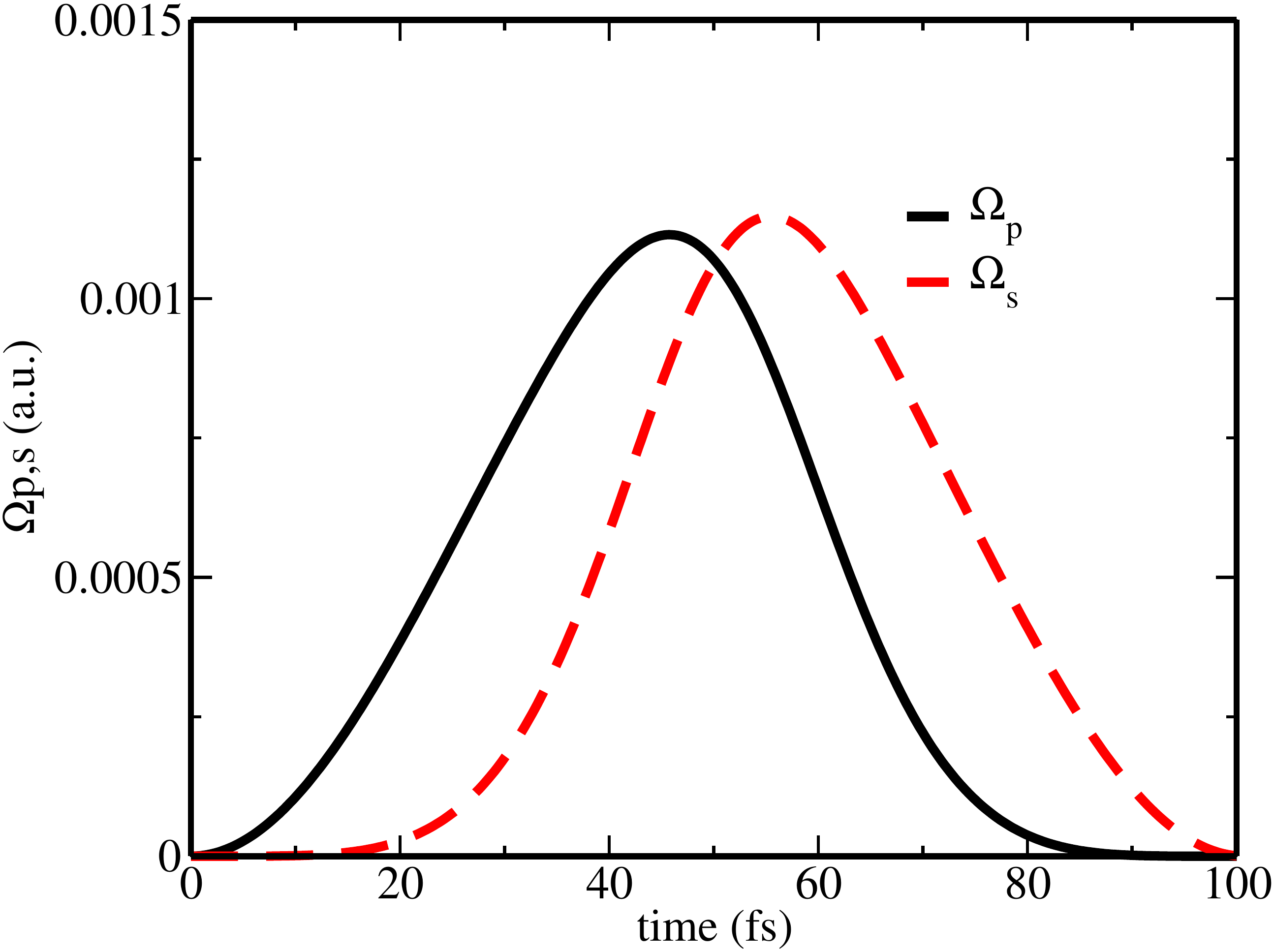}
\caption{The sequence of the two optimized pulses obtained 
using the standard OCT method. We note that the intuitive sequence 
of pump pulse (solid dark curve ) followed by Stokes pulse (red dashed curve) 
is generated automatically  
\label{fig:Pulses-OCT-Standard}}
\end{figure}

In such case the functional $J$ is written as the following: 
\begin{eqnarray} 
  J_{\mathrm{st}} & = & |\langle\varPsi_i(T)|\varphi_f\rangle|^2 -\int_0^T 
\lambda_a 
{(E(t)-E_\mathrm{ref})^2 dt}, 
\label{Eq:Fonctional}
\end{eqnarray}
where the index $st$ refers to the standard OCT method. $E(t)$ 
denotes the electric field and $E_\mathrm{ref}$ is the reference field.  
The optimization could be carried out only within a specific time frame [0,T]. 
The 
first term in Eq.~(\ref{Eq:Fonctional}) represents the objective or the yield. 
It serves to measure the quality of the control, it may, inter alia, 
overlaps with the target state $|\varphi_f\rangle$ (present case) or determine 
an expected 
value for an hermitian operator. The role of the second term is to limit the 
laser energy through the penalty factor $\lambda_a(t)=\lambda_0/S(t)$, where 
$S(t)=\sin^2(\pi t/T)$. It constraints the intensity of the laser to not exceed 
a certain limit, called the Keldysh limit \cite{Keldysh}. 
\\
Maximizing the objective functional Eq.~(\ref{Eq:Fonctional}) can be achieved 
by  
resolving numerically a system of three coupled equations \cite{Rabitz1}: 
The Schr\"{o}dinger equation for $|\varPsi_i(t)\rangle$ with initial condition 
$|\varPsi_i(t=0)\rangle=|\varphi_i\rangle$ (conventionally called forward 
propagation), the Schr\"{o}dinger equation for the Lagrange multiplier with a 
final condition $|\varPsi_f(t=T)\rangle=|\varphi_f\rangle$ (conventionally 
called backward propagation), in addition to an equation for the optimal field. 
The system of equations to be resolved can be written as the following:  
 \begin{subequations}
 \begin{eqnarray}
   \frac{\partial}{\partial t} |\varPsi_f(t)\rangle 
   &=& -i \hat{H}|\varPsi_f(t)\rangle,  
    \, |\varPsi_f(T)\rangle =  |\varphi_f\rangle
   \,, 
   \label{eq:system1}
   \\
    \frac{\partial}{\partial t} |\varPsi_i(t)\rangle &=& -i
   \hat{H} |\varPsi_i(t)\rangle,  
    \qquad |\varPsi_i(t = 0)\rangle = |\varphi_i\rangle\,, 
   \label{eq:system2}
    \\
    E(t) &=& E_{\mathrm{ref}}(t) + \Delta E(t)
   \label{eq:system3}
    \\
    \Delta E(t) &= & \frac{S(t)}{\lambda_a} \Im \left(\langle \varPsi_f(t)| 
  \hat{\mu} |\varPsi_i(t)\rangle \right). \nonumber
 \end{eqnarray}
   \label{eq:system}
 \end{subequations}
Where $\hat\mu$ represent the dipole coupling.
In order to resolve these coupled equations different methods have been 
proposed \cite{Rabitz1, Palao2} in the past. For instance, the Krotov method 
involves 
rather an 
iterative procedure which, simultaneously, leads to the maximization of the 
functional at the end of the process. We therefore use Krotov method in our 
study. In 
fact, this method needs to adjust the field over the time towards a monotonic 
convergence in a self-consistent way. Further details on how Krotov method 
operates could be found in \cite{Tannor2, Somloi, Palao2}.
Our purpose here is to demonstrate that the standard OCT method cannot lead to 
a proper coherent 
population transfer in the three-level system shown in Fig. 
\ref{fig:lambda_system}. 
For a same pulse duration $T=100$ fs as in STIRAP example, we display the  
populations evolution as a function of time in 
Fig.~\ref{fig:Populations-OCT-Standard}. We can see in 
Fig.~\ref{fig:Populations-OCT-Standard}(a) that standard OCT leads to 
a transfer of population from state $|1\rangle$ represented here in 
green dashed line to state $|3\rangle$ represented by the solid black 
line. However this transfer could not be coherent, where we notice in 
Fig.~\ref{fig:Populations-OCT-Standard} (b) a big 
population amount had to go through the intermediate state $|2\rangle$ 
represented by the dotted red line. We notice that around $\simeq50\%$ of the 
population reside in state $|2\rangle$ at $t=50$ fs.
Figure.~\ref{fig:Pulses-OCT-Standard} shows the natural intuitive feature 
that occurs when we adopt the standard OCT method. The pump pulse 
presented in solid dark curve is clearly followed by the Stokes pulse presented 
in red dashed curve. In fact, this behavior is the expected one since as 
initial 
conditions we apply the pump pulse before the Stokes.
\subsection{OCT with State-Dependent Constraint}
In reference \cite{Tannor} it was clearly mentioned that  ``{\it Without 
robustness 
being incorporated explicitly into the objective functional in OCT there is no 
reason to expect STIRAP-type solutions to emerge from OCT calculation}''. 
Here we point out that the state-dependent constraint included into the 
objective functional \cite{Koch} leads to this kind of particular solution 
where the counterintuitive scheme is plainly visible. 

By introducing the state-dependent constraint within the objective 
functional Eq.~(\ref{Eq:Fonctional}), one could obtain:

\begin{eqnarray} 
  J_{\mathrm{new}} & = & J_{\mathrm{st}} +\underbrace{\lambda_b\int_0^T 
\langle\varPsi_i(t)|\hat{D}|\varPsi_i(t)\rangle
dt}_{\mathrm{state-constraint}};
\label{Eq:Fonctional_new}
\end{eqnarray}
with $\hat{D} =|1\rangle\langle 1| + |3\rangle\langle 3|$. The second term of 
Eq.~(\ref{Eq:Fonctional_new}) describes the state-dependent constraint. 
The objective of the optimization herein consists of avoiding population 
transfer to level $|2\rangle$. 
As defined by Palao et al. \cite{Koch}, here the states $|1\rangle$ and 
$|3\rangle$ form  the allowed subspace. Contrariwise state $|2\rangle$  
corresponds to the forbidden subspace.
The optimization can be achieved by solving the 
three couple equations Eq.~(\ref{eq:system}). However the equation of backward 
propagation, Eq.~(\ref{eq:system1}), should be modified in order take into 
account the state-dependent constraint. Therefore, we obtain the following 
equation which has to be resolved along with other equations of the system by 
using the same technique adopted in standard OCT method. 
 \begin{equation}
   \frac{\partial}{\partial t} |\varPsi_f(t)\rangle 
   = -i \hat{H}|\varPsi_f(t)\rangle
   +\lambda_b \hat{D}|\varphi(t)\rangle,  
    \, |\varPsi_f(T)\rangle =  \varphi_f\rangle.
    \label{Eq:EDP-new}
 \end{equation}

Eq.~(\ref{Eq:EDP-new}) is solved numerically by using a Chebychev propagator 
for inhomogeneous Schr\"odinger equation \cite{Ndong}.
\begin{figure}[h!]
\centering
\includegraphics[width=0.7\linewidth]{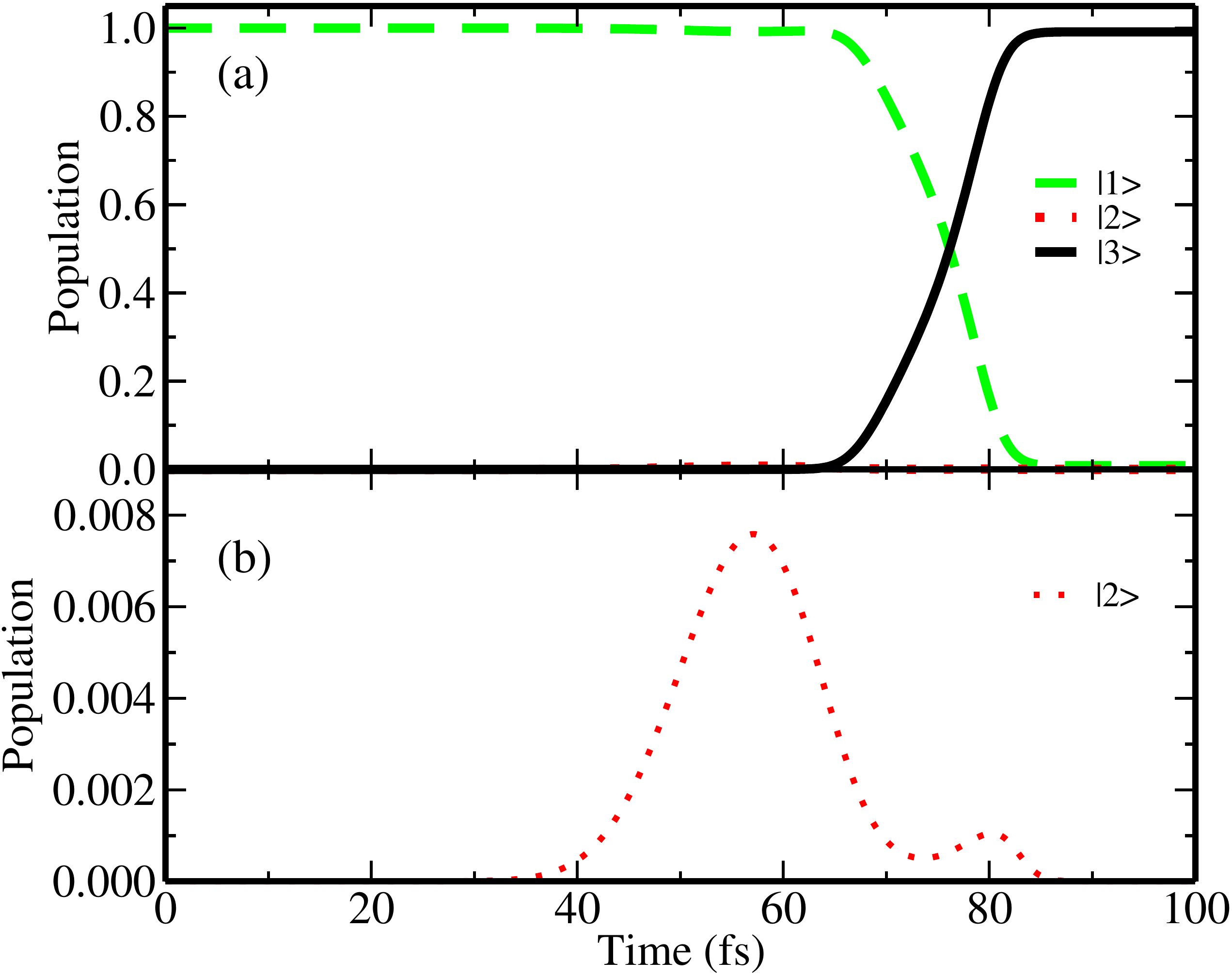}
\caption{The time evolution of the population in the three states 
$|1\rangle>$ (green dashed line), $|2\rangle>$ (red dotted line), and 
$|3\rangle>$ (dark solid line) with laser parameters generated by the new 
OCT method with state-dependent constraint. 
\label{Fig:Populations-OCT-New}}
\end{figure}
Figure \ref{Fig:Populations-OCT-New}(a) shows the time evolution of the 
population in the three-level system generated by the new OCT method with the 
state-dependent constraint. One can observe a better population transfer from 
level $|1\rangle$ presented in green dashed line to level $|3\rangle$ presented 
in dark solid line. 
In order to examine the efficiency of this transfer, the 
population in the intermediate level $|2\rangle$ which has been locked in 
the objective functional is shown in Fig \ref{Fig:Populations-OCT-New}(b). One 
could see that the time evolution of population in level $\left|2\right>$ 
attains its maximum of $0.75\%$ at $t=59$ fs. Then the population 
completely vanished at the end of the pulse. This result reveals to be more 
interesting compared to the result obtained from the local 
optimization. One could notice in Fig.7 of \cite{Bartana} that the population 
in level $|2\rangle$ has a higher value around $1\%$ and keeps 
have the same over its time evolution until the pulse is off. 
In addition compared to the standard OCT method, the population transfer is 
enhanced significantly indeed. The amount of the population went through the 
level $|2\rangle$ with the standard OCT method was around $50\%$, and yet we 
have less than $0.75\%$ in same level with the new OCT method.

\begin{figure}[h!]
\centering
\includegraphics[width=0.7\linewidth]{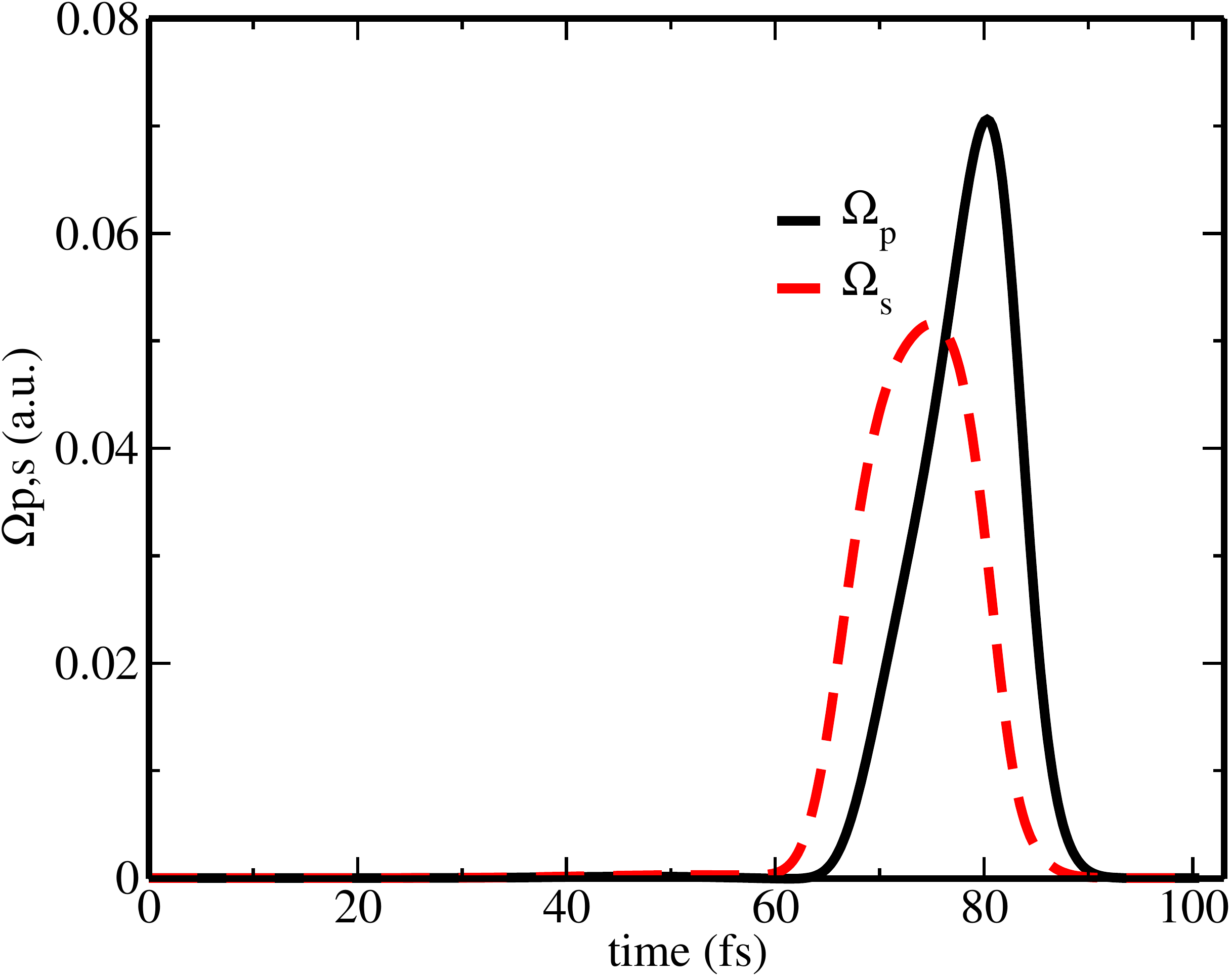}
\caption{The sequence of the two optimized pulses obtained 
using the new optimal control theory method with state-dependent constraint. We 
distinctly see the counterintuitive sequence of Stokes pulse (red 
dashed line) followed by pump pulse (solid dark line) is generated 
automatically as in STIRAP method. 
\label{Fig:Pulses-OCT-NEW}}
\end{figure}

Furthermore, in Fig.~\ref{Fig:Pulses-OCT-NEW} we display the optimized pulses 
obtained by the new OCT method. Although, we do not 
obtain the same envelope for both pulses, we see fairly that the Stokes 
presented in dashed red line precedes and overlaps the pump presented in 
solid dark line. We thus retrieve the counterintuitive feature of STIRAP 
in a systematic manner. 
\subsection{Comparison of Robustness}
So far we have shown that the new OCT technique can produce the 
counterintuitive scheme as in STIRAP method. Here, we analyze the robustness 
of these two method with respect to the pulse fluctuations and the decay of the 
intermediate state $|2\rangle$. 
Firstly, we introduce fluctuations in the optimized pluses by adopting 
the following technique:
\begin{equation}
  \Omega_{s,p}'(t) = \Omega_{s,p}(t) +\alpha \zeta(t),
\end{equation}
where $\Omega_{s,p}(t)$ is the solution obtained by the new OCT  or 
STIRAP. Indexes $s$ and $p$ refer respectively to the Stokes and 
pump pulses. $\zeta(t)$ is the perturbation function modeled by a vector of 
random numbers chosen within the interval $[-1, 1]$ and $\alpha$ is a factor 
parameter. 
\\ 
Secondly, in order to investigate the robustness of the two methods with 
respect to the decay, we introduce a simple loss mechanism in level 
$\left|2\right>$. 
For the obvious reason that the goal of STIRAP scheme in a three-level system 
is to transfer the population from level $|1\rangle$ to level 
$|3\rangle$ (see Fig.~\ref{fig:lambda_system}) by minimizing the population 
of the intermediate level $|2\rangle$. This loss mechanism could be 
modeled by adding an imaginary term $-i\beta\Gamma$ to the energy of the level 
$\left|2\right>$, where $\Gamma = 1/{T}$ represents the decay rate
and $\beta$ is a second factor parameter. 

Consequently the corresponding new Hamiltonian becomes:
 \begin{eqnarray}
   H' &=& 
     \begin{bmatrix}
        0 &  \Omega_p'(t) &  0 \\
       \Omega_p'(t) & \Delta -i\beta\Gamma &   \Omega_s'(t) \\
       0 &  \Omega'_s(t) &  0 
     \end{bmatrix}.
     \label{eq:Ham2}
   \end{eqnarray}
  
We then solve the Schr\"odinger equation using the new Hamiltonian given in 
Eq.~(\ref{eq:Ham2}). We analyze separately the robustness, firstly with respect 
to the pulse fluctuations and secondly with respect to the decay. 
In our analysis, the amplitude of the Rabi frequencies of the STIRAP approach 
is 
chosen in a way that the condition for adiabatic following
is fulfilled. This condition is given in Ref. \cite{Bergmann} by 
$\Omega_{s,p}^{\mathrm{max}}\tau\gg 10$
with $\tau$ the time overlap of the two Rabi frequencies and
$\Omega_{s,p}^{\mathrm{max}}$ the maximum of the Rabi frequency.  
\begin{figure}[h!]
\centering
\includegraphics[width=0.7\linewidth]{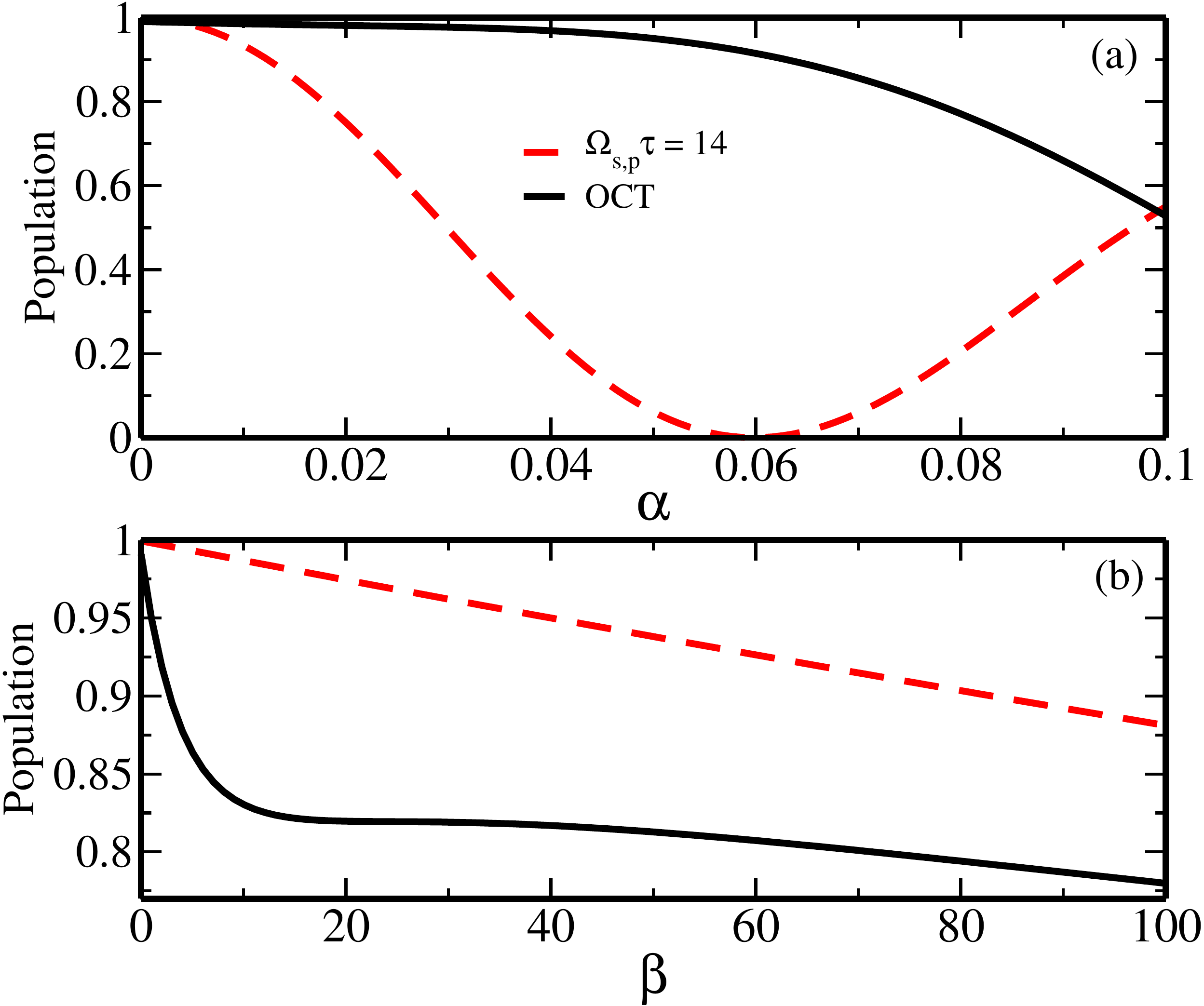}
\caption{Population at final time in the target level $|3\rangle$ induced 
by the Hamiltonian of Eq.~(\ref{eq:Ham2}). 
In the upper panel, $\beta$ parameter in  Eq.~(\ref{eq:Ham2}) is set to 0 
while in the lower panel $\alpha=0$,.
The dashed red curve shows the results obtained with the STIRAP method and the 
solid black curve represents the results 
obtained with the new OCT approach.}
\label{fig:pop_robustness_energy}
\end{figure}
The upper panel of Fig.~\ref{fig:pop_robustness_energy} compares the robustness
between the STIRAP and the new OCT approach with respect to the pulse 
fluctuations. While the lower panel of  Fig.~\ref{fig:pop_robustness_energy} 
compares the robustness of these two methods with respect 
to the decay introduced in Eq.~(\ref{eq:Ham2}). 
The figure displays the population at the final instant of the 
target level $|3\rangle$  as a function of $\alpha$ parameter 
(Fig.~\ref{fig:pop_robustness_energy}(a)) 
or $\beta$ parameter (Fig.~\ref{fig:pop_robustness_energy}(b)).
In addition to the adiabatic following condition, 
for the results shown in Fig.~\ref{fig:pop_robustness_energy}, we ensured that 
the energy of the STIRAP field is set to be equivalent to the energy of the 
optimized field generated by the new OCT method.  
Since in the STIRAP approach are defined analytically, one expect that 
this method would be more robust. But surprisingly, it is clearly seen from the 
solid black line of the upper panel of 
Fig.~\ref{fig:pop_robustness_energy} that the new OCT method is less 
sensitive to the field perturbation than the STIRAP approach.  
Once a field perturbation is introduced, the dashed red line that describes the 
results obtained with the STIRAP decreases drastically. For example for 
$\alpha=0.06$, the population in the target state is only about 2\% for the 
STIRAP method while it is greater than 90\% for new OCT method. The fact that 
the optimized pulses are more robust than the ones produced by STIRAP with 
regards to the pulse fluctuations is attributed to the pulses duration. In fact 
the FWHM of the optimized pulses is two times shorter than the STIRAP pulses. 

Contrariwise in the lower panel of Fig.~\ref{fig:pop_robustness_energy} we 
observe an opposite behavior. The STIRAP method appears to be 
more robust with respect to the decay. The dashed red curve shows a linear 
decrease of the population. For instance, when $\beta=100$, 
the population in the target state is about 87\% for the STIRAP method. The
solid black curve which displays the results obtained with new OCT method shows 
an exponential decrease when $0\leq\beta\leq10$. At $\beta=100$, the population 
in the target state is about 78\% for the new OCT method.
Hence, at first glance, one could conclude that the robustness of each of these 
methods depends on what we take into consideration which is usually dictated by 
the experiment setup.
\begin{figure}[h!]
\centering
\includegraphics[width=0.7\linewidth]{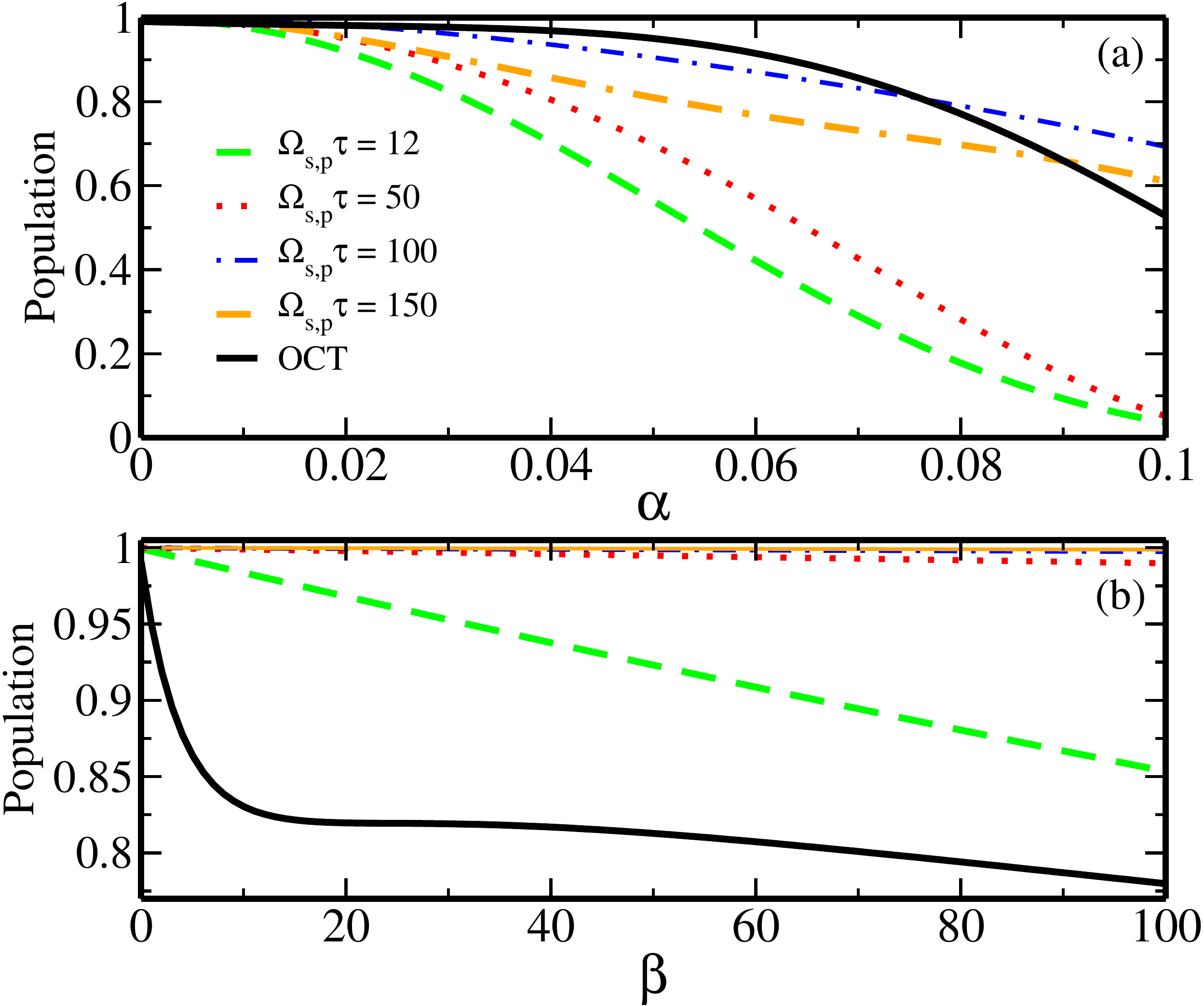}
\caption{ Same as in Fig.~\ref{fig:pop_robustness_energy}.
Comparison of robustness for several field energies of the STIRAP method.}
\label{fig:pop_robustness}
\end{figure}
In order to investigate more the robustness and going further in 
respecting the adiabatic following condition 
$\Omega_{s,p}^{\mathrm{max}}\tau\gg 
10$, we show in figure \ref{fig:pop_robustness} the same comparison as in
Fig.~\ref{fig:pop_robustness_energy} the population of the 
target state when the pulse is off, yet for different field energies of the 
STIRAP pulse. In panel (a), the results corresponding to $\Omega\tau=100$ and 
$\Omega\tau=150$ show that the robustness of the STIRAP with respect to the 
pulse fluctuations does not necessarily increase when we increase the field 
energy as one could expect. However the efficiency with regard to the 
decay is significantly improved when we increase the energy of the initial 
pulse, Fig.~\ref{fig:pop_robustness_energy}(a). Yet we 
should emphasize on two points here, regarding the efficiency as a function of 
the decay, indeed STIRAP appears to be more robust, however the new OCT 
still keep having high population within the target state, the minimum value is 
78\% when $\beta=100$, in addition to the fact that we did not increase the 
energy of the initial pulse for OCT. The second point is about the adiabatic 
following condition, it is of course very important to fulfill this condition 
to ensure a population transfer within STIRAP. However it implies to choose 
higher energies for the initial pulses, which usually represents a kind of 
burden for experimental setups.
This analysis allows us to conclude that, for the population transfer from an
initial state to a final one, the OCT method with the 
sate-dependent constraint is more robust than the STIRAP when perturbations are
introduced in the field. However STIRAP is more efficient when we refer to the 
decay in the intermediate
state of the system.  

\subsection{Many-level system }
\begin{figure}
\centering
\includegraphics[width=0.5\linewidth]{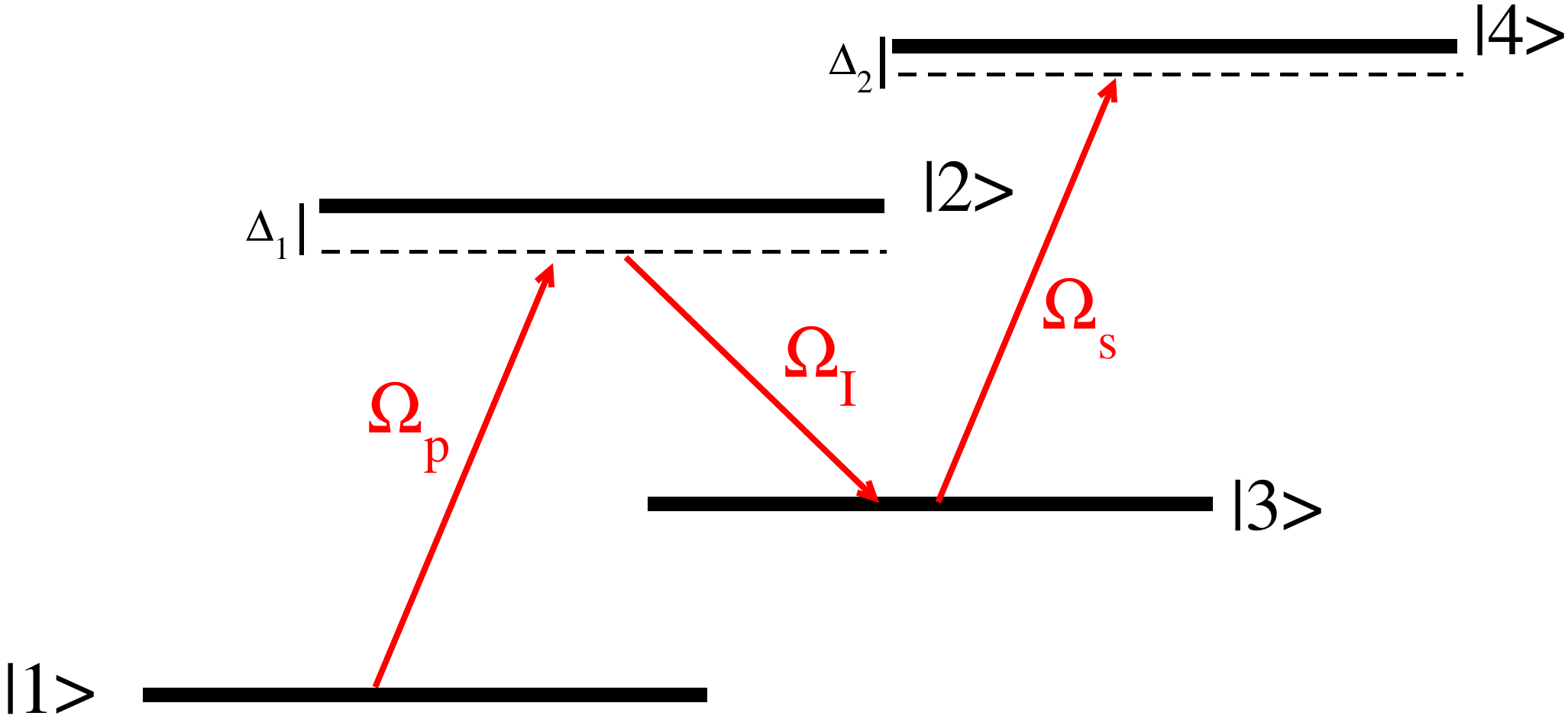}
\caption{A four-level system with sequential couplings. Level $|1\rangle$ and 
$|2\rangle$ are coupled by a field with amplitude $\Omega_p$. Levels 
$|2\rangle$ and $|3\rangle$ are coupled by a field with amplitude $\Omega_I$ 
and levels $|3\rangle$ and $|4\rangle$ are coupled by a field with amplitude 
$\Omega_s$. $\Delta_1$ and $\Delta_2$ are respectively the detuning of levels 
$|2\rangle$ and $|4\rangle$. 
\label{Fig:four_level}}
\end{figure}

\begin{figure}[h!]
\centering
\includegraphics[width=0.7\linewidth]{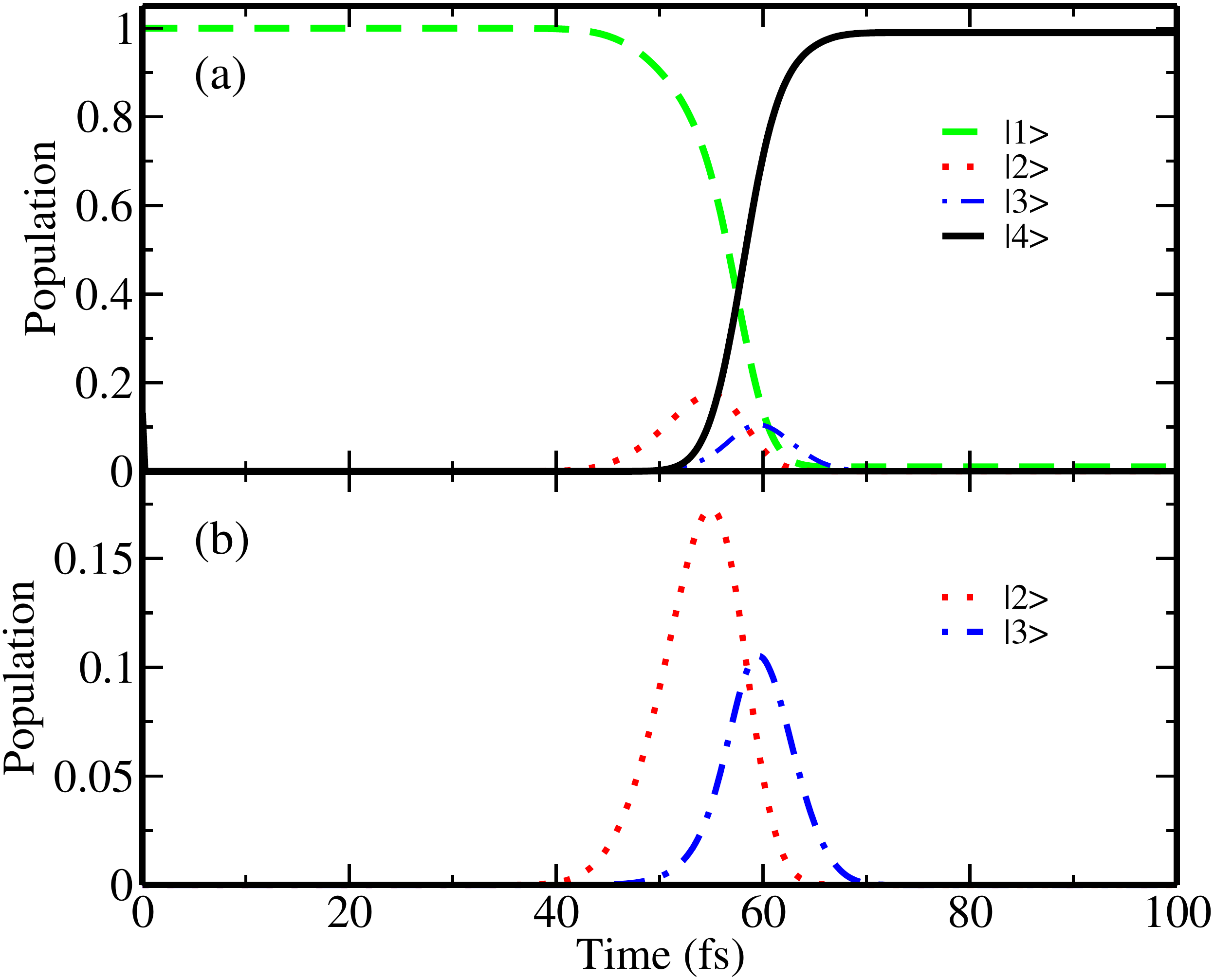}
\caption{The population evolution in a four level 
system with a sequential coupling. Panel (a) shows the population transfer from 
the initial state $|1\rangle$ (green dashed curve) to the final 
state $|4\rangle$ (solid black line). Panel (b) is displaying the population 
evolution of the intermediate states $|2\rangle$ (red dotted curve) 
and $|3\rangle$ (blue dot-dashed curve).
\label{Fig:pop_four_octnew}}
\end{figure}

\begin{figure}[h!]
\centering
\includegraphics[width=0.7\linewidth]{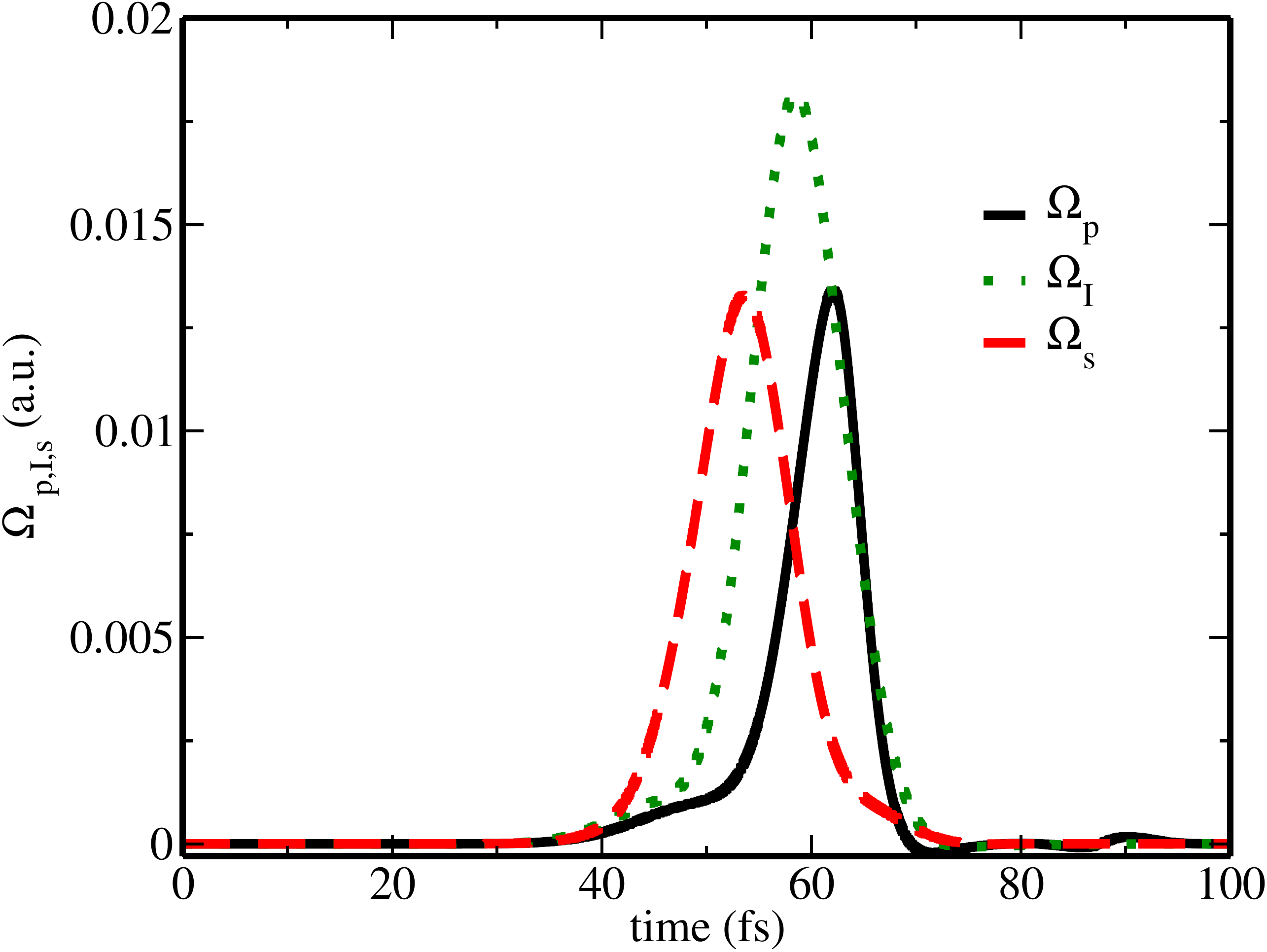}
\caption{The sequence of the optimized pulses in a sequentially coupled 
four-level system. The pump pulse is presented in solid black line, the Stokes 
in dashed red line. The intermediate pulse which couple the two intermediate 
levels $|2\rangle$ and $|3\rangle$ is presented by the dotted green line. Note 
here that the counterintuitive sequence of the pump pulse preceded by the 
Stokes is generated automatically. We remark also the intermediate pulse 
overlaps the two other pulses and its intensity is much higher comparing to the 
intensities of the Stokes and the pump. This phenomena was obtained in the past 
by using LCT calculation and was called "straddling" STIRAP.
\label{Fig:pulse_four_octnew}}
\end{figure}

One of the major results of reference \cite{Tannor} is the general 
applicability of LCT. It was shown that LCT can be straightforward implemented 
for N-level systems to reproduce STIRAP scheme. By using a STIRAP pulses 
sequence, LCT led to a population transfer within four- and nine-level systems.
We show here that OCT with the state-dependent constraint can be extended to 
multilevel systems as well, and lead to a complete population transfer. For the 
sake of simplicity, we focus here on a four-level system where the coupling 
between the different levels is rather sequential as shown in 
Fig~\ref{Fig:four_level}. 
\skip 5.0 in
\skip 2.0 in
In Fig.~\ref{Fig:pop_four_octnew}(a), we present the population transfer in the 
case of four-level system. In similar fashion, the process of the transfer from 
the initial state $|1\rangle$, presented in dashed green curve, to the target 
state $|4\rangle$, presented in solid black curve, is analogue to the one in 
the $\Lambda$-type three-level system. 
We obtain indeed an almost complete population transfer. In 
Fig~\ref{Fig:pop_four_octnew}(b), we focus on the two intermediate states 
$|2\rangle$ (dotted red line) and $|3\rangle$ (dot-dashed blue line) in order 
to examine the quality of the transfer. One can see that the 
percentages of population passed through these two states remain very small. 
Only around $15\%$ through state $|2\rangle$ at $t=$55 fs and less 
than $10\%$ through state $|3\rangle$ at $t=$ 60 fs, we refer evidently here to 
the two peaks in Fig~\ref{Fig:pop_four_octnew}(b).
The optimized pulses generated from the four-level system by using the new OCT 
method are presented in Fig.~\ref{Fig:pulse_four_octnew}. One should remind 
that in this case we need three pulses; the pump to link the level $|1\rangle$ 
to level $|2\rangle$, the Stokes to link level $|3\rangle$to level $|4\rangle$ 
and the third pulse which we call the intermediate pulse to link the two 
intermediate levels $|2\rangle$ and $|3\rangle$. 
As in the three-level system, we obtain automatically the counterintuitive 
mechanism. One can see clearly the pump pulse (solid black line) precedes the 
Stokes pulse (dashed red line). Nevertheless we observe the intermediate pulse 
$\Omega_I$ overlaps both the Stokes and the pump pulses. This phenomena is 
quite equivalent to the one obtained from the local optimization \cite{Tannor}, 
where indeed the envelope of the intermediate pulse straddle the Stokes and the 
pump, yet with a higher intensity.
\skip 0.2in
In reference \cite{Tannor}, authors attributed to this pattern the name of 
\textit{stradding} STIRAP sequence (S-STIRAP) and demonstrated that is a 
robust extension of STIRAP to multilevel systems. They illustrated indeed the 
applicability of this method to five- and nine-level systems. Since the new 
OCT method leads to equivalent results for three- and four-level systems, we 
expect that it would conduct to similar conclusion for N-level systems.  

\section{CONCLUSION}
With reference to Tannor's work \cite{Tannor} we have shown that 
STIRAP-type solution can be generated from an OCT calculation. To obtain this 
solution, a state-dependent constraint has been added to the objective 
functional. Comparisons with STIRAP in a $\Lambda$-type three-level 
system have shown that this new OCT formulation leads to equivalent 
population transfers with a minimum of loss in forbidden levels for same 
initial 
energies. In addition, it has proven itself to generate slightly better 
population transfer when we have pulse fluctuations introduced into the system. 
However STIRAP appears to be slightly more robust with respect to the decay 
of the intermediate forbidden state. 
Furthermore the advantage of using this new method is the fact that it could be 
extended from a three-level system to multilevel systems.   
\section*{Acknowledgment(s)}
The authors would like to thank Prof. Christiane Koch her helpful 
discussions.


\section*{References}
\begin{thebibliography}{99}
\bibitem{Marlan}
Marlan O. Scully and M. Suhail Zubairy, Quantum Optics (Cambridge University 
Press, New York, (1997).
\bibitem{Bergmann}
K. Bergmann, H. Theuer and B. W. Shore, Reviews of Modern
Physics 70 (1998) 1003.
\bibitem{Bergmann2}
U. Gaubatz, P. Rudecki, S. Schiemann and K. Bergmann, Journal of Chemical 
Physics 92 (1990) 5363. 
\bibitem{Vitanov}
N. V. Vitanov, T. Halfmann, B. Shore, K. Bergmann, Annual Review of Physical 
Chemistry 52 (2001) 763.
\bibitem{Fewell}
M. P. Fewell, B. W. Shore and K. Bergmann, Australian Journal of Physics 50 
(1997) 281.
\bibitem{Fleischhauer}
M. Fleischhauer and Aaron S. Manka, Physical Review A 54 (1996) 794.
\bibitem{Kobrak}
M. N. Kobrak and S. A. Rice, Physical Review A 57 (1998) 1158.
\bibitem{Band}
Y. B. Band and O. Magner, Physical Review A 50 (1994) 584.
\bibitem{Elk}
M. Elk, Physical Review A 52 (1995) 4017.
\bibitem{Schiemann}
S. Schiemann, A. Kuhn, S. Steuerwald and K. Bergmann, Physical Review Letters 
71 (1993) 3637.
\bibitem{Jaouadi}
A. Jaouadi, E. Barrez, Y. Justum  and M. Desouter-Lecomte, Journal of Chemical 
Physics 139 (2103) 014310
\bibitem{Kuklinski}
J. R. Kuklinski, U. Gaubatz, F. T. Hioe and K. Bergmann, Physical Review A 40 
(1989) 6741.
\bibitem{Guerin}
S. Gu\'erin, H. R. Jauslin, Advances in Chemical Physics 125 (2003) 147.
\bibitem{Tannor}
V. S. Malinovsky and D. J. Tannor,  Physical Review A 56 (1997) 4929.
\bibitem{Bartana}
D. J. Tannor, R. Kosloff and A. Bartana, Faraday Discussions 113 (1999) 
365.
\bibitem{Yuan}
H. Yuan, C. P. Koch, P. Salamon and D. J. Tannor, Physical Review A 85 (2012) 
033417.
\bibitem{Koch}
J. P. Palao, R. Kosloff and C. P. Koch,  Physical Review A 77 (2008) 
063412.
\bibitem{Muller}
M. M. M\"uller, D. M. Reich, M. Murphy, H. Yuan, J. Vala, K. B. Whaley, T. 
Calarco and C. P. Koch, Physical Review A 84 (2011) 042315.
\bibitem{Rabitz}
Y. Ohtsuki, G. Turicini, H. Rabitz, Journal of Chemical Physics 120 (2004) 5509.
\bibitem{Vivie}
K. Sundermann, R. de Vivie-Riedle, Journal of Chemical Physics 110 (1999) 1896.
\bibitem{Keldysh}
M. V. Ammosov, N. B. Delone and V. P. Krainov, Soviet Physics. Journal of 
Experimental and Theoretical Physics 64 (1986) 1191.
\bibitem{Rabitz1}
W. Zhu, J. Botina, H. Rabitz, Journal of Chemical Physics 108 (1998) 1953.
\bibitem{Palao2}
J. P. Palao and R. Kosloff, Physical Review A 68 (2003) 062308.
\bibitem{Ndong}
M. Ndong, H. Tal-Ezer, R. Kosloff, C. P. Koch, Journal of Chemical Physics 130 
(2009) 124108.
\bibitem{Tannor2}
D. J. Tannor, V. Kazakov, and V. Orlov, Time Dependent Quantum Molecular 
Dynamics (1992) 347.
\bibitem{Somloi}
J. Somloi, V. A. Kazakov and D. J. Tannor, Chemical Physics 172 (1993) 85.

\end{thebibliography}
\end{document}